\documentstyle[preprint,aps]{revtex}

\preprint{
\font\fortssbx=cmssbx10 scaled \magstep2
\hbox to \hsize{
\hbox{\fortssbx University of Wisconsin - Madison}
\hfill\vbox{\hbox{\bf MADPH-95-888}
                 \hbox{\bf astro-ph/9512080}
                \hbox{December 1995}}}}

\begin{document}

\title{\vspace*{.5in}
Ultra-Transparent Antarctic Ice as a Supernova Detector}

\author{F.~Halzen$^1$, J.E.~Jacobsen$^1$ and E.~Zas$^2$}

\address{
$^1$Department of Physics, University of Wisconsin, Madison, WI 53706, USA\\
$^2$Dpto.\ de Part\'\i culas, Universidad de Santiago, E-15706 Santiago, Spain}

\maketitle

\begin{abstract}
We have simulated the response of a high energy neutrino telescope in deep
Antarctic ice to the stream of low energy neutrinos produced by a supernova.
The passage of a large flux of MeV-energy neutrinos during a period of seconds
will be detected as an excess of single counting rates in all individual
optical modules. We update here a previous estimate of the performance of such
an instrument taking into account the recent discovery of absorption lengths of
several hundred meters for near-UV photons in natural deep ice. The existing
AMANDA detector can, even by the most conservative estimates, act as a galactic
supernova watch.
\end{abstract}

\newpage

\section{Introduction}

Although aspects of the observations of SN1987A\cite{Kamioka,IMB} left some
lingering doubts about supernova models\cite{LSD,IMBang}, they provided, in
general, remarkable confirmation of established ideas about supernova
mechanisms\cite{mechanism}. At collapse, the core of the progenitor star is
expected to release energy in a prompt
$\nu_e$ burst lasting a few milliseconds. Most of the energy is, however,
liberated after deleptonization in a burst lasting about ten seconds. Roughly
equal energies are carried by each neutrino species with a thermal spectrum of
temperature $2-4~$MeV. The time scale corresponds to the thermalization of the
``neutrinosphere'' and its diffusion within the dense
core\cite{sneutrinos}.

Since the ${\bar{\nu}}_e$ cross-section\cite{xsection} on protons in the
detector is significantly larger than the interaction cross sections for the
other neutrino flavors, ${\bar{\nu}}_e$ events dominate the signal by a large
factor after detection efficiency is taken into account. In this reaction, free
protons absorb the antineutrino to produce a neutron and a positron  which is
approximately isotropically emitted with an energy close to that of  the
initial neutrino. A thermal spectrum of temperature $4~$MeV, when folded with
an inverse beta decay cross section which increases with the square of the
neutrino energy, yields an observed positron energy distribution which peaks in
the vicinity of $20~$MeV.

High energy neutrino telescopes under construction deploy Optical Modules
(OM's) in a clear medium which acts as the radiator of \v Cerenkov light from
muon tracks or electromagnetic showers produced in neutrino interactions. They
are primarily designed to exploit the long range of high energy muons and have
typical nominal threshold energies in the GeV-range, ostensibly too high to
observe supernova-neutrinos. However, the interactions of  $\bar\nu_e$ with
protons produces copious numbers of positrons with tens of MeVs of energy.
These will yield signals in all OMs during the (typically 10 second) duration
of the burst. Such a signal, even if statistically weak in a single OM, will
become significant for a sufficient number of OMs. The number of OMs required
for monitoring galactic supernovae has been shown to be of order a few hundred,
a number typical for the first-generation detectors under
construction\cite{halzen}. Apart from the excess signal and its time profile,
additional information on the energy, direction or neutrino species is unlikely
to be reconstructed in such telescopes, in contrast to dedicated supernova
experiments.

The AMANDA-detector, operating in low background, sterile Antarctic ice, is at
an advantage as far as detection of supernova bursts is concerned. The
background counting rates in the OMs are dictated by the dark current of the
photomultiplier only. They are reduced by over an order of magnitude compared
to natural sea-water because of the absence of bioluminescence and radioactive
decay of potassium\cite{Webster}. Recently, a second advantage of using polar
ice has been revealed. With the first calibration measurements it has become
evident that the absorption length of light for in-situ South Pole ice is
larger than what had been anticipated\cite{askebjer,ICRC}. The absorption
length of deep South Pole ice has the astonishingly large value of $\sim310$~m
for the 350 to 400~nm light to which the PMTs are sensitive. A value of only
8~m had been anticipated from laboratory measurements. For supernovae neutrino
signals the detector volume scales linearly with the absorption length. This
can be demonstrated from a simple back-of-the-envelope estimate of the
effective detection volume of an OM for MeV-positrons.

The track-length of a 20~MeV positron in ice is roughly 12 centimeters and
therefore over 3000 \v Cerenkov photons are produced. This number combined
with a typical quantum efficiency of $25~\%$ leaves $N_{\gamma} \simeq 800$
detectable photons in each event. At large
distances the probability that the OM detects a photon falls as $R^{-2}$
\begin{equation}
P(R) \simeq (R_d / R)^2 \,\,\,\,\,\,\,\, {\rm for} \,\, R>>R_d
\end{equation}
Here $R$ is the distance to the positron shower. $R_d$ is defined such that
photon detection probability by the OM is essentially unity for distances
smaller than $R_d$, and falls quadratically for larger distances. Its value can
be estimated from the relation
\begin{equation}
\pi R_d^2 \sim N_{\gamma} A,
\end{equation}
where $A$ is the photocathode area of the OM. The relation states that the
effective area for guaranteed detection is proportional to the brightness of
the source and to the photocathode area. We next calculate the effective volume
$V_{\rm eff}$ associated with each OM by integrating the probability function
over  volume:
\begin{equation}
V_{\rm eff} \simeq \int_0^{R_{\rm abs}} P (R) R^2 dR \,.
\end{equation}
The integral is, of course, cut off by the absorption length $R_{\rm abs}$ of
the  light. It is easy to see that its value is dominated by $R > R_d$ and
negligible for $R < R_d$. Therefore
\begin{equation}
V_{\rm eff}  \simeq \int_{R_d}^{R_{\rm abs}} P (R) R^2 dR
\sim {1 \over \pi} N_{\gamma} A R_{\rm abs}\,,
\end{equation}
for $R_{\rm abs}\gg R_d$.
This estimate can be made more quantitative\cite{halzen}, but is sufficient to
demonstrate that the effective volume for supernova positrons,
corresponding to each OM, is
proportional to both the collection area of the OM and the attenuation length
of the light. We will demonstrate that the discovery of very large values of
$R_{\rm abs}$ for natural polar ice implies that the AMANDA detector can
function
as a galactic supernova watch.

At present, four AMANDA strings have been deployed in Antarctic ice at the
South Pole.  Each string holds 20 OMs separated by 10 meters at depths between
800 and 1000 meters. The calibration measurements have revealed absorption
lengths of light for in-situ ice which are larger than what had been
anticipated from laboratory experiments\cite{askebjer,ICRC}. The results have
been obtained by studying the propagation of \v Cerenkov light from high energy
muons, and by studying the propagation of laser light pulsed from small nylon
spheres attached to the ends of an optical fiber below each OM. The results
indicate that the propagation of photons in the ice is
consistent with a diffusive process with very large absorption lengths. The
diffusive character of the process apparently results from scattering of the
light by residual bubbles trapped in the ice. The number density of such
bubbles is observed to decrease with depth. Direct measurements on ice cores
indicate that such bubbles totally disappear at 1--1.5~km depth, depending on
location. Clearly these results have important implications for supernova
neutrino detection because we expect a linear enhancement of the effective
volume per OM with absorption length.

We have calculated the effective volume of the present AMANDA detector for
supernova neutrinos, taking into account the increased attenuation length,
and have also considered the effect of scattering of the light by residual air
bubbles at 1~km depth. The above back-of-the-envelope estimate yields an
effective volume $V_{\rm eff} \sim 2000$~m$^3$ for an AMANDA OM with collecting
area $A=0.028$~m$^2$  and an absorption length $R_{\rm abs} \simeq 300$~m. The
effective radius of each  OM for detecting supernova neutrinos is about
7~meters. This is very  encouraging and indicates that a single OM is similar
in collection volume to  the Kamiokande and IMB detectors.

We have confirmed and sharpened this result with a complete Monte Carlo
simulation. Positron showers were generated using a 4~MeV Boltzman
distribution multiplied by $E^2$ in order to represent the increased detection
efficiency associated with the rise of the neutrino interaction cross section
with energy\cite{xsection}. We used an electromagnetic shower Monte Carlo
which was originally designed to calculate radio pulses in ice and correctly
describes showers down to MeV energies\cite{zasemp}. The output, in the form of
\v Cerenkov photon distributions radiated by positron showers, was fed into a
simulation of the AMANDA detector. This program treated in detail both the
propagation of the photons, their absorption and scattering, and the
efficiencies of the OMs. For the current AMANDA configuration, with 73 stable
OMs, the simulations reveal an effective volume per OM $V_{\rm eff} \simeq 1.7
\pm 0.4~10^3~$m$^3$, compatible with our
estimate and more than an order of magnitude above the results previously
obtained assuming the properties of laboratory ice\cite{halzen}. The observable
signal is totally dominated by events where a positron shower yields a single
photoelectron hit in a single OM. The quoted error is statistical. Although the
absorption length at the peak acceptance of the OMs is well measured, its
dependence on color in the near UV region is poorly determined. More precise
future measurements may increase or decrease our result but cannot modify it
significantly.

We have verified by Monte Carlo that the effect of residual bubbles on photon
propagation is within our statistical error. This is to be expected because our
results imply that the photons are collected over a radius of 7.4~meters, a
distance much shorter than the absorption length of the light. For a scattering
length $R_s$ of order 0.25~meters\cite{askebjer,ICRC} photons within this
radius will reach the PMT whether scattered or not. The scattering is
irrelevant as long as the actual distance travelled by the scattered photons
($N R_s$, with N the number of times that the photon scatters), is shorter than
$R_{\rm abs}$. For a random walk the number of scatters $N$ for a photon
originating a distance d from the OM is given by
\begin{equation}
N \simeq d^2 / R_s^2
\end{equation}
The proportionality factor depends on the angular distribution of the
scattering process; it is 4/3 in the present problem\cite{askebjer,ICRC}. The
condition that the pathlength $N R_s < R_{\rm abs}$ becomes
\begin{equation}
d < \sqrt{{4 \over 3} R_s R_{\rm abs}}
\end{equation}
which is satisfied for all distances $d$ up to 7.4~meters for $R_s \simeq
0.25$~m and $R_{\rm abs} \simeq 310$~m.

These results can be used to rescale SN1987 observations to a supernova at a
distance of $d_{\rm kpc}$.  To rescale the data from the Kamiokande(IMB)
experiment  to the AMANDA detector we need the ratio of their effective volumes
corrected  for the different thresholds of the instruments. This introduces a
correction  factor $f$ which is evaluated from the Fermi-Dirac energy
distribution and the  $E^2$-dependence of the cross section. Using a supernova
temperature for  SN1987A of $4.0$~MeV\cite{mechanism} $f$ is about 0.8 (0.2)
for Kamiokande  (IMB) because of the reduced threshold of AMANDA operating as a
pure counting  detector. From the 11 events observed in the 2.14~kton
Kamiokande detector we  predict:

\begin{equation}
N_{\rm Events} \sim 11~N_M~\left[{\rho~V_{\rm eff} \over f~2.14~{\rm kton}}
\right]
\left[ {52~{\rm kpc} \over d_{\rm kpc}} \right]^2
\end{equation}
For the present AMANDA configuration with $N_M=73$, $\rho=0.924$~g~cm$^{-3}$,
the density of ice, and
$V_{\rm eff}=1700$~m$^3$, we obtain 270 events per OM for a total of nearly
20,000  events.

We must now require a meaningful detection of this signal in the presence of
the continuous background counting rate of all phototubes. Over the 10~seconds
duration of the neutrino burst, the RMS-fluctuations of the combined noise
from all the OMs is:
\begin{equation}
\sigma_{1p.e.} = {\sqrt {(10\rm~sec) \nu_{1p.e.}~N_M }}.
\end{equation}
Here $\nu_{1p.e.}$ represents the average background counting rate in a single
module at the 1 photoelectron level, and equals1.86~kHz for the AMANDA OMs. The
probability that this noise fakes a supernova signal can be, at least
theoretically, estimated from Poisson
statistics. The expected rate of supernova explosions in our galaxy is about
$2\times10^{-2}$~y$^{-1}$. If the detector is to perform as a supernova watch
we must require that the frequency of fake signals is well below this rate.
The signal should therefore exceed $n_{\sigma} \ge 6$ which corresponds to a
probability of $9.9\times10^{-10}$. The corresponding number of 10~second
intervals does indeed exceed a century. This requirement can be relaxed if
we just demand that the detector can make a measurement in the presence of
independent confirmation. This may be achieved by operating parts of a single
instrument as independent detectors.

For an average noise rate of 1.86~kHz the RMS fluctuation of the 1.36 million
hits expected in an interval of ten seconds is 1165. This yields a detected
excess of 17~$\sigma$ for a supernova at the center of the galaxy. Demanding a
6~$\sigma$ signal for a supernova
watch, the present AMANDA configuration with appropriate triggering, would view
a radius of 17~kpc, thus covering our galaxy.
Clearly signal-to-noise is proportional to $\sqrt{N_M}$ and therefore the
number of OMs needed for fixed signal-to-noise ratio scales with the distance
to the fourth power. A next-generation detector with 6000 OM's would provide a
sharp signal for a supernova within a sphere of radius 52~kpc.

The results are extremely encouraging and an effort should be made to find the
optimal trigger. In a realistic detector the background
counting rate does not follow a Poisson distribution with a width given by
simple
statistics. The actual performance of the present AMANDA data acquisition
relevant to the detection of supernovae is described elsewhere\cite{Ralph}. The
AMANDA detector, frozen in then Antarctic ice, can act for decades as a
supernova watch with minimal maintenance or investment in manpower and
operation of the instrument.

\acknowledgements

This work was supported in part by the University of Wisconsin
Research Committee with funds granted by the Wisconsin Alumni Research
Foundation, in part by the U.S.~Department of Energy under Grant
No.~DE-FG02-95ER40896, in part by the Xunta de Galicia under
contract XUGA-20604A93 and in part by and the CICYT under contract AEN93-0729.

\end{document}